\documentstyle[aps,graphicx,epsf,latexsym]{revtex} 
\makeatletter
\def\thepage{\@arabic\c@page}
\def\@pnumwidth{2em}
\def\REVTeX{REV\TeX}
\makeatother
\begin{document}
\twocolumn

\setlength{\unitlength}{1cm}

\title{Phase Behavior of a Simple Model for Membrane Proteins}
\author{Massimo G. Noro} 
\address{
Unilever Research Port Sunlight\\
Quarry Road East, Bebington Wirral CH63 3JW, U.K.}
\author{Daan Frenkel}
\address{
FOM Institute for Atomic and Molecular Physics\\
Kruislaan 407, 1098 SJ Amsterdam, The Netherlands}
\date{\today}
\maketitle

\makeatletter
\global\@specialpagefalse
\def\@oddhead{Typeset with \REVTeX{} \hfill Preprint}
\let\@evenhead\@oddhead
\def\@oddfoot{\reset@font\rm\hfill \thepage\hfill
\ifnum\c@page=1
  \llap{\protect\copyright{}
  Submitted to The Journal of Chemical Physics}%
\fi
} \let\@evenfoot\@oddfoot
\makeatother

\begin{abstract}
We report a numerical simulation of the phase diagram of a simple model for
membrane proteins constrained to move in a plane. In analogy with the
corresponding three dimensional models, the liquid-gas transition becomes metastable
as the range of attraction
decreases. Spontaneous
crystallization happens much more readily in the two dimensional models
rather than in their three dimensional counterparts.
\end{abstract}

\section{Introduction}

If one picture is worth a thousand words, recent advances in X-ray
crystallography are providing the equivalent of a dictionary.
Crystallographers are now solving the three-dimensional structure of
proteins at the rate of one or more per day. A bottleneck is the difficulty
of growing high-quality crystals for X-ray analysis.

As the success of protein crystallization depends strongly on the physical
conditions of the initial solution\cite{McPherson,Durbin}, much effort has
gone into finding the relation between solvent conditions and
crystallization behavior. Rosenbaum {\it et al.} \cite{RZZ} analysed the
solubility curves of a variety of globular proteins and found that they can
be made to superimpose when expressed in appropriate scaled units. What this
suggests is that the phase behavior of many globular-protein solutions obeys
a law of corresponding states. Specifically, they showed that the
solid-fluid phase boundary of the proteins in solutions can be mapped onto
the corresponding curve of a simple model system (the hard-core Yukawa
potential\cite{Hagen}) with short-ranged attractions. \ Such
corresponding states behavior suggests that --- for a given class of compounds
--- the solubility boundary is only weakly dependent on the details of the
interaction potential.

Far fewer crystals have been grown of membrane proteins than of globular
proteins. The reason is simply that it is more difficult to crystallize
membrane proteins than globular proteins. It would, of course, be very
interesting to know if the ``generic'' features of the phase behavior of
quasi-two-dimensional proteins are similar to those of globular proteins. It
is tempting to start such an analysis by looking at the corresponding
``minimal'' model for membrane proteins --- namely one of circular disks with
isotropic, short-ranged attractive interactions. In this paper, we report a
numerical study of the phase behavior of such model membrane proteins. In
our study, we vary the range of the attractive interaction between the
particles. We find that the general topology of the phase diagram is indeed
similar to that for three-dimensional (``globular'') proteins. In
particular, we find that the liquid-vapor transition is preempted by the
freezing curve for particles with sufficiently short-ranged attractions.
However, quantitatively, there are large differences. In contrast to the
three-dimensional case which still exhibits a well-defined meta-stable
liquid-vapor coexistence curve below the equilibrium freezing curve, we find
crystallization, rather than fluid-fluid demixing, in the corresponding
``membrane-protein'' model.

\section{Model}

\begin{figure}[tbp]
\centering{
\begin{minipage}{6.7cm}
    \epsfxsize 6.0 cm
    \rotatebox{-90}{\epsfbox{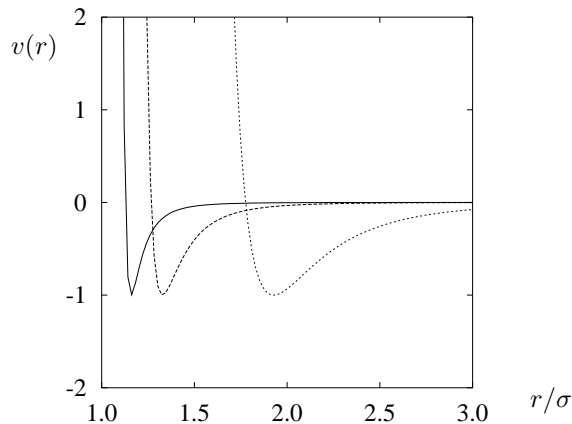}}
  \end{minipage}
\put(-7.1, 2.2){$v(r)$} 
\put(-0.2,-2.5){$r/\sigma$}}
\par
\caption{Interaction potential as defined in Eq.\ref{Potential}
corresponding to $\protect\alpha=50$ 
(solid line),
$\protect\alpha=4.5$ 
(dashed line),
and $\protect\alpha=0.1$ 
(dotted line).
Note that as $\protect\alpha$
decreases, attractions become longer ranged but the range of repulsions
increases as well, so that the ``effective'' size of the particle grows.
(We have set $\epsilon=1$).}
\label{Potentials}
\end{figure}

We model the effective interaction between membrane proteins using an
extension of the well-known Lennard-Jones potential, 
\begin{equation}
v(r)=\left\{ \matrix{ \infty \hfill \cr \frac{4\epsilon}{\alpha^2}
\left(\frac{1}{\left[(r/\sigma)^2-1\right]^6}
-\frac{\alpha}{\left[(r/\sigma)^2-1\right]^3}\right) \hfill \cr }\right.
\quad \matrix{ r\le \sigma \hfill \cr \sigma <r \hfill \cr }\qquad ,
\label{Potential}
\end{equation}
where $\sigma $ denotes the hard-core diameter of the particles and $%
\epsilon $ the well depth. The width of the attractive well can be adjusted
by varying the parameter $\alpha $: the smaller $\alpha $ the longer the
range of attractions. Figure (\ref{Potentials}) plots this potential for the
values of $\alpha $ used in this paper. Note that as $\alpha $ decreases,
the range of repulsions increases as well, so that the ``effective'' size of
the particle grows. It is, however, convenient to compare the simulation
results for particles that have the same hard-core diameter. To estimate the
effective hard-core diameter for a given value of $\alpha $, we separate the
potential into an attractive $v_{att}$ and a repulsive $v_{rep}$
contribution in the spirit of the Weeks-Chandler-Andersen method\cite
{Andersen}. We then calculate the equivalent hard-core size of the repulsive
part of the potential using the Barker-Henderson criterion\cite{Barker}:

\begin{equation}
\sigma _{eff}=\int_{0}^{\infty }dr\left[ 1-e^{v_{rep}(r)/k_{B}T}\right]
\qquad .  \label{SigEff}
\end{equation}
In what follows, we use $\sigma _{eff}$ as our unit of length. In these
units, all our model proteins have the same effective diameter.
%

\section{Computational Methods}

\label{Method}In order to map out the phase diagram of our two-dimensional
model for membrane proteins, we have used a combination of simulation
techniques that we discuss briefly in this section.

\subsection{Gibbs-Duhem integration}

This method was proposed by Kofke\cite{Kofke.1,Kofke.2}, and is based on the
integration of the Clausius-Clapeyron equation, which expresses the slope of
the phase boundary in the ($P$,$\beta $) diagram, 
\begin{equation}
\frac{dP}{d\beta }=-\frac{\Delta h}{\beta \Delta v}  \label{CC}
\end{equation}
where $\beta =1/k_{B}T$, $P$ is the pressure, $\Delta h$ and $\Delta v$ are
the differences in enthalpy and volume (per particle) in the two coexisting
phases, respectively. To compute this slope, two simulations are carried out
in parallel: one in the liquid phase and one in the solid. The two systems
are held at the same temperature and pressure, but cannot interact with each
other; during the runs we measure the average density ($1/v$) and enthalpy $%
h $ per particle and thus determine $(dP/d\beta )$. Knowing this slope, we
can then estimate the location of \ a neighboring point on the ($P$,$\beta $%
) coexistence curve. The Gibbs-Duhem method is straightforward, but there
are several ways in which it can be implemented, and there are several
subtleties that require attention.

The first is the choice of a good starting point. The Gibbs-Duhem
integration method allows one to trace out the ($P$,$\beta $) coexistence
curve once one point on this curve is known. There are several ways to
select this point. In one set of calculations, we started the step-wise
integration at $\beta =0$ ({\bf i.e.} the infinite temperature limit) where the
phase diagram approaches that of hard disks. Here we have used the 
fluid-solid equilibrium density
gap reported by Jaster\cite{Jaster} as the input of our first Gibbs-Duhem
simulation. However, as the freezing transition of hard disks is itself
still not completely characterized \cite{Bates}, this is not necessarily the
best option. In fact independent free-energy calculations allow us to start
the Gibbs-Duhem integration at a finite value of $\beta $ where we find a
strong first-order liquid-solid phase transition. We found this second route more
reliable.

\begin{figure}[tbp]
\begin{minipage}{6.7cm}
    \epsfxsize 6.0 cm
    \rotatebox{-90}{\epsfbox{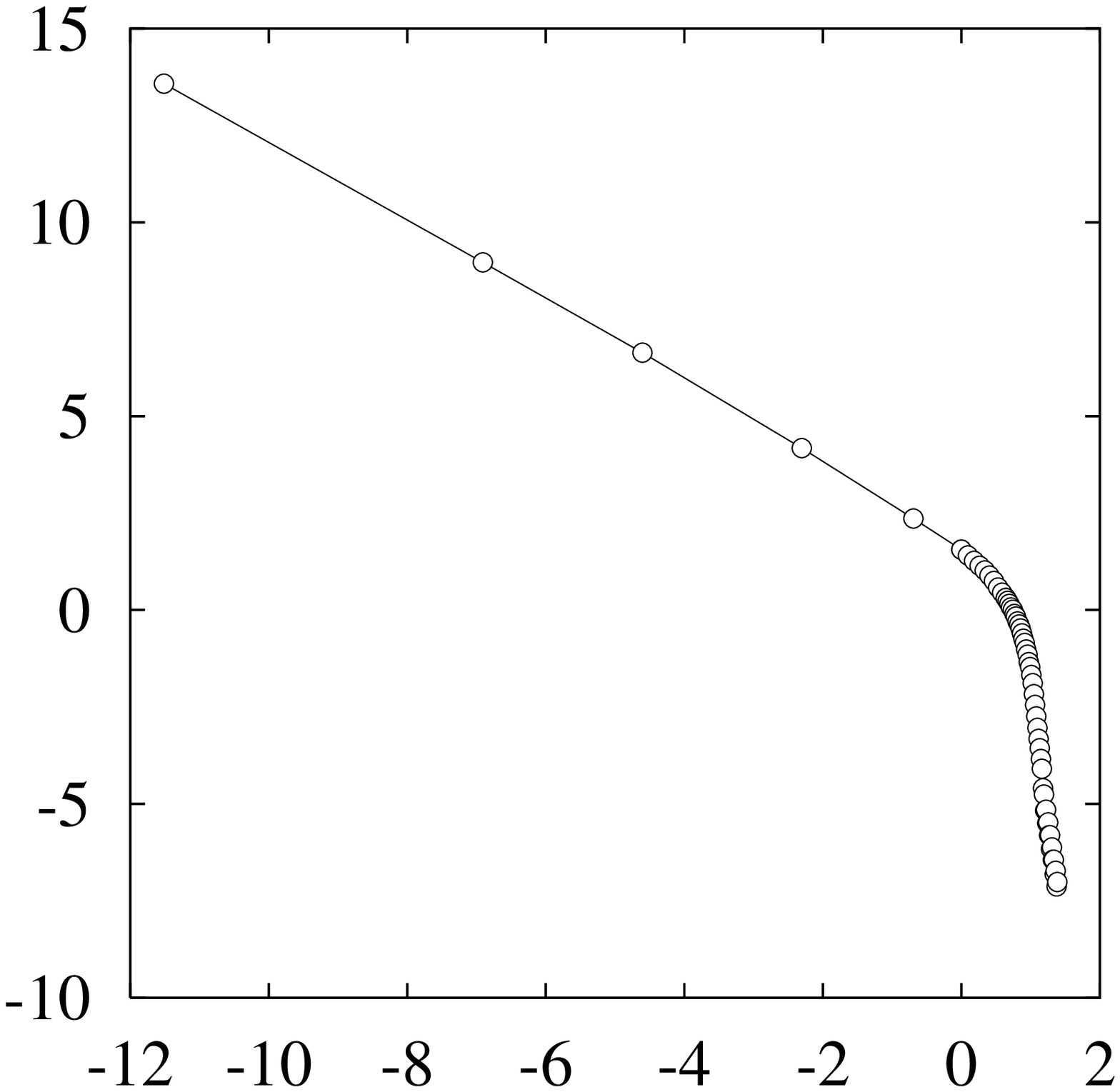}}
    \center{(a)  ($\log P$,$\log \beta$) integration}
  \end{minipage}
\put(-7.1, 3.2){$\log P$} 
\put(-0.2,-2.0){$\log \beta$}
\par
\vspace{1cm}
\begin{minipage}{6.7cm}
    \epsfxsize 6.0 cm
    \rotatebox{-90}{\epsfbox{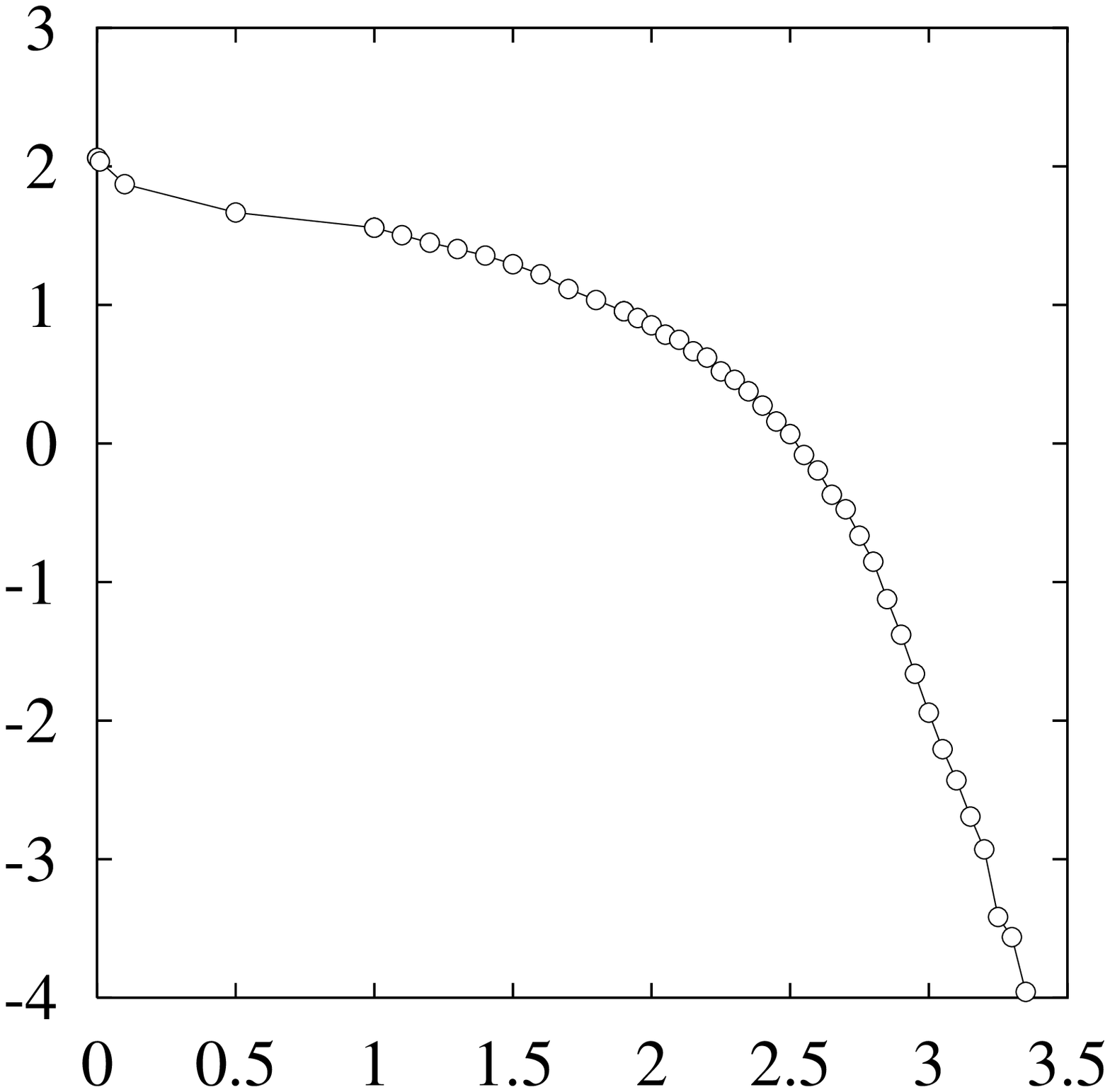}} 
    \center{(b)  ($\log \beta P$,$\beta$) integration}
  \end{minipage}
\put(-7.4, 3.2){$\log \beta P$} 
\put(-0.2,-2.0){$\beta$} 
\par
\caption{Phase boundary in the two representations used for the Gibbs-Duhem
integration scheme: (a) in the high temperature regime (b) in the low
temperature regime.}
\label{PhaseBoundary}
\end{figure}

The second point is the choice of the form of the equation to be integrated.
The Gibbs-Duhem method is not self-correcting. This means that small
numerical errors may cause the computed coexistence curve to diverge from
the true phase-equilibrium curve. To minimize this problem, the right-hand
side of Eq. \ref{CC} must be a smooth function of pressure and temperature,
so that simple integration schemes can be applied with high accuracy. In our
case we found that at high temperatures a suitable slowly varying function
was 
\begin{equation}
\frac{d\log P}{d\log \beta }=-\frac{\Delta e+P\Delta v}{P\Delta v}
\label{CClog}
\end{equation}
where $\Delta e$ is the difference in the energy per particle between the
two coexisting phases. In the hard-particle limit the $P\Delta v$ term
completely overwhelms the energy difference, and the slope of the phase
boundary plotted in the ($\log P$,$\log \beta $) diagram approaches $-1$.
We have verified this in our simulations. In the low temperature regime
the most convenient differential form of the Clausius-Clapeyron equation, in
agreement with Kofke\cite{Kofke.1} and Hagen\cite{Hagen}, was found to be: 
\begin{equation}
\frac{d\log \beta P}{d\beta }=-\frac{\Delta e}{\beta P\Delta v}\qquad .
\label{CCen}
\end{equation}
Equations \ref{CClog} and \ref{CCen} were solved using a second-order
predictor-corrector algorithm. As such algorithms are not self-starting, we
initiated the integration by supplying the values of the integrand and its
slope (when known), and using a first-order algorithm for the prediction of
the first point; after two integration steps we continued with the desired
second order procedure. In Figure (\ref{PhaseBoundary}) we collect the phase
transition points obtained from numerical integration of 
equations \ref{CClog} and \ref{CCen} .

The third point has to do with spontaneous phase transitions during a single
phase simulation. In two dimensions there is much less hysteresis in the
solid-liquid transition than in three dimensions. As a consequence, it may happen 
that, in
a constant-pressure simulation of a relatively small system (in our case: $%
N=256$), a fluid could {\it spontaneously} transform into a solid, or vice
versa. This creates a problem for the Gibbs-Duhem simulations that involve
constant-pressure studies of state points along the solid-fluid coexistence
line. To prevent such undesirable (and, on a macroscopic scale, irrelevant)
fluctuations, we imposed a constraint on the degree of crystallinity of the
system. The degree of crystallinity was measured using a global bond-order
parameter\cite{Steinhardt}. If during a constant-pressure Monte Carlo
simulation of the liquid (or solid) phase, the value of the bond-order
parameter of a configuration was outside the interval typical for the phase
under consideration, then the configuration was rejected. Typically no more
than 0.5\% of the configurations were rejected during a simulation of any
state point.
%

\subsection{Thermodynamic Integration}

\begin{figure}[tbp]
\centering{
\begin{minipage}{6.7cm}
    \epsfxsize 6.0 cm
    \rotatebox{-90}{\epsfbox{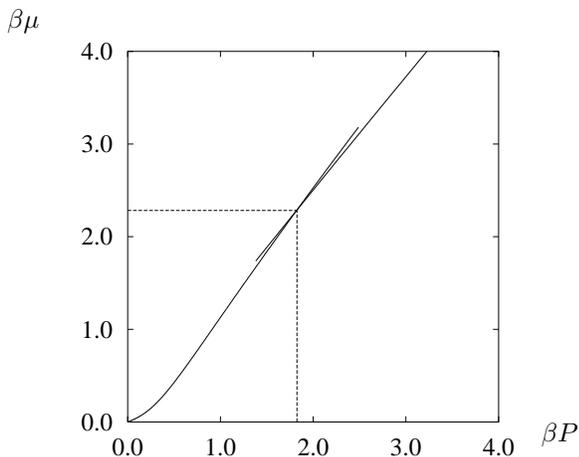}}
  \end{minipage}
\put(-7.3, 3.0){$\beta \mu$} 
\put(-0.2,-2.5){$\beta P$}}
\par
\caption{Plot of the chemical potential $\protect\mu$ as a function of the
pressure $P$ for the liquid and the solid phase. The transition condition is
satisfied at the crossing: at a fixed temperature the chemical potentials
and the pressure are equal. Here we neglect the De Broglie wavelength
contribution. The calculated equilibrium pressure does not change since one
subtracts the same quantity, namely $\ln \protect\lambda^2$, from the
chemical potential of both phases. }
\label{mu-P}
\end{figure}

\begin{figure}[tbp]
\begin{minipage}{6.7cm}
    \epsfxsize 6.0 cm
    \rotatebox{-90}{\epsfbox{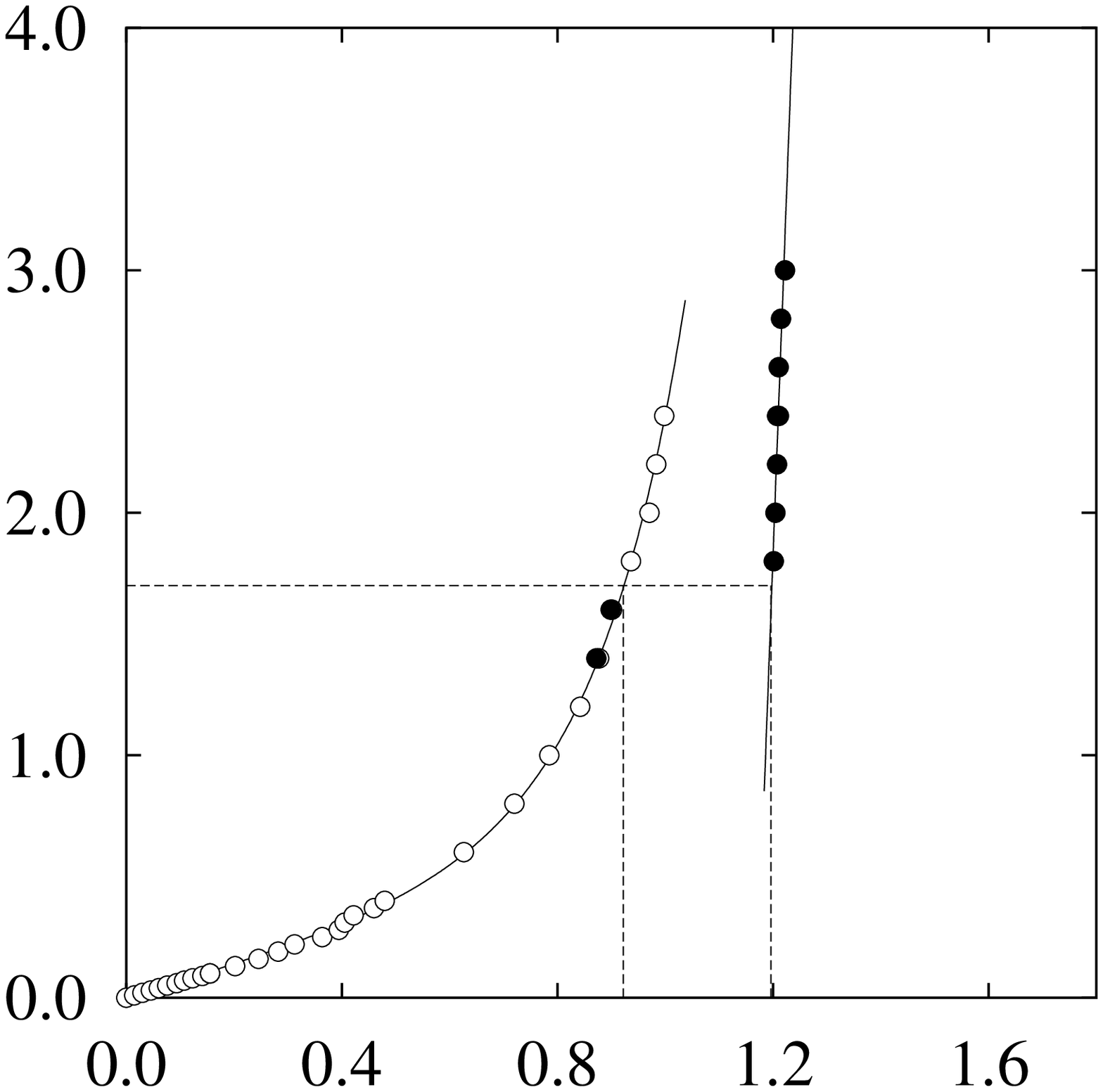}}
    \center{(a) $\alpha=50 \qquad \beta=2$}
  \end{minipage}
\put(-7.1, 3.2){$\beta P$} 
\put(-0.2,-2.0){$\rho \sigma_{eff}^3$}
\par
\vspace{1cm}
\begin{minipage}{6.7cm}
    \epsfxsize 6.0 cm
    \rotatebox{-90}{\epsfbox{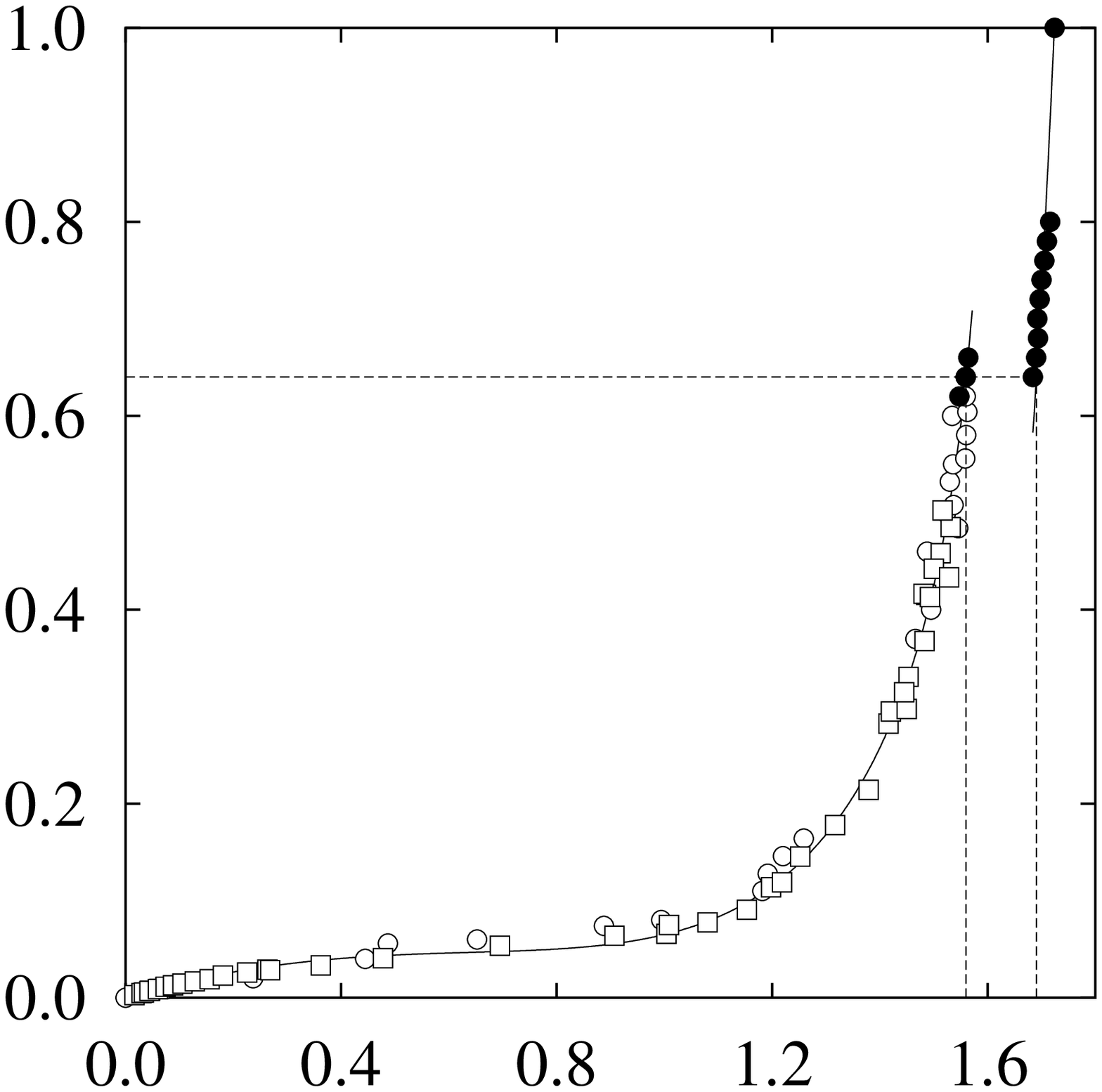}} 
    \center{(b) $\alpha=0.1 \qquad \beta=2$}
  \end{minipage}
\put(-7.1, 3.2){$\beta$} 
\put(-0.2,-2.0){$\rho \sigma_{eff}^3$} 
\par
\caption{Equation of state, or ($P$,$\protect\rho$)-isotherm, calculated at
the same temperature for two different ranges of interactions: (a) estremely
short ($\alpha=50$), (b) extremely long ($\alpha=0.1$). The simulation data have 
been collected by:
measuring the average density in NPT-Monte Carlo simulations when
compressing a liquid 
(open circles)
or when expanding a solid 
(filled circles),
and 
(open squares)
measuring the average pressure in $\protect\mu$VT-Monte Carlo simulations.}
\label{P-rho}
\end{figure}

Thermodynamic integration is the method most commonly used to locate the
solid-liquid transition. The procedure involves the comparison of the
chemical potentials of the liquid and solid phases at equal temperature and
pressure. The result of such a calculation is a ($\mu $,$P$)-diagram similar
to the one shown in Figure (\ref{mu-P}). Two phases are in equilibrium
at the point where
the two chemical potential curves cross. Evaluation of the chemical
potential of the fluid branch is straightforward once the equation of state
of the fluid is known from low densities up to the density range of
interest. In the present case, we did this by performing a large number of
NPT-Monte Carlo simulations at different state points
(see Figure (\ref{P-rho})). We then fitted the
numerical data for the pressure to a convenient fitting function. In the
present case, we used an ad-hoc generalization of the so-called 
$y$-expansion that is often used to describe the equation of state of hard-body
fluids\cite{Gelbart}: 
\begin{equation}
\beta P=\frac{\rho }{1-a\rho }+b\left( \frac{\rho }{1-a\rho }\right)
^{2}+c\left( \frac{\rho }{1-a\rho }\right) ^{3}\qquad ,  \label{Yexp}
\end{equation}
where $a$, $b$ and $c$ are to be determined from the fit. Upon integration
of the pressure (\ref{Yexp}) between zero density (ideal gas limit) and the
density of interest, one obtains an explicit expression for the chemical
potential: 
\begin{eqnarray*}
\beta \mu (\rho )=
\ln \frac{\rho \Lambda ^{2}}{1-a\rho }
+\frac{b/a-c/a^{2}+1}{1-a\rho }
+\frac{c/2a^{2}+b\rho }{\left( 1-a\rho \right) ^{2}}\\
+\frac{c\rho^{2}}{\left( 1-a\rho \right) ^{3}}
-\left( b/a-c/2a^{2}+1\right) 
\end{eqnarray*}
\begin{equation}
\label{ExChemY}
\end{equation}
where $\Lambda $ is the De Broglie wavelength. In the case of the solid
phase, we fit the calculated ($P$,$\rho $)-isotherms to a simple power law
of the form $a\rho ^{2}+b\rho +c$. Integrating between a reference density $%
\rho ^{\ast }$ and the present density yields the chemical potential 
\begin{eqnarray*}
\beta \mu (\rho )=
2a\rho 
+b\left( \ln \rho +1\right) 
-\left( a\rho ^{\ast}
+b\ln \rho ^{\ast }
-c/\rho ^{\ast }\right)\\
+\beta f^{\ast ex}(\rho ^{\ast})
+\ln \Lambda ^{2}\rho ^{\ast }
-1
\qquad ,  
\end{eqnarray*}
\begin{equation}
\label{ExChemPL}
\end{equation}
where $f^{\ast ex}(\rho ^{\ast })$ is the excess Helmholtz free energy per
particle evaluated at the reference density $\rho ^{\ast }$. A method that
is widely used to compute the free energy of a crystalline solid is the so
called ``Einstein crystal'' method proposed by Frenkel and Ladd\cite
{Frenkel-Ladd}, which employs thermodynamic integration of the Helmholtz
free energy along a reversible artificial path between the system of
interest and an Einstein crystal. The Einstein system is used as a reference
since there is a simple analytical expression for its partition function,
which allows a determination of the {\it absolute} value of its free energy.
We typically performed a series of 10 NVT-simulations at the reference
density $\rho ^{\ast }$, switching gradually from the Einstein crystal to
our system of interest, by modifying the value of the coupling constant $%
\lambda $, where $\lambda =1$ corresponds to the Einstein crystal and $%
\lambda =0$ to the system of interest. The value of the excess free energy
takes a simple form\cite{Polson}: 
\begin{eqnarray*}
\beta f^{\ast ex}(\rho ^{\ast })=
-\frac{d}{2}\ln \left[ \frac{2\pi }{\alpha\beta }\right] 
+\frac{\beta U_{0}}{N}
-\frac{\beta }{N}\int_{0}^{1}d\lambda\left\langle \Delta U\right\rangle _{\lambda }\\
-\frac{N-1}{N}\ln \rho +1
-\frac{d+1}{2N}\ln N
-\frac{1}{2N}\ln 2\pi 
\qquad , 
\end{eqnarray*}
\begin{equation} 
\label{RefSol}
\end{equation}
where $d$ is the dimensionality, $N$ is the size of the system, $\alpha $
is the spring constant of the Einstein crystal, and $U_{0}$ the potential
energy of the crystal with all the atoms in their lattice positions. 
The difference between the energy of the Einstein crystal and that of the 
system of 
interest, $\Delta U\equiv U_{Ein}-U$, enters in the third term in Eq. \ref{RefSol} 
and the integral in this term is evaluated numerically as explained in 
Ref.\cite{Frenkel-Smit}.
%

Once the transition pressure is known, from the crossing of the two 
chemical potential curves
of Figure (\ref{mu-P}), one simply reads off the coexisting densities from
the ($P$,$\rho $) diagram. In Figure (\ref{P-rho}) we show two typical
equations of state for the case $\alpha =50$ and $\alpha =0.1$ where we have
collected state points using both NPT-simulations and $\mu $VT-simulations (%
{\it i.e.} performed in the Grand Canonical Ensemble). At low temperatures
it becomes increasingly difficult to equilibrate the fluid system,
especially if the series of simulations is performed as a gradual
compression. The occasional formation of high density clusters of particles
generates locally a highly incompressible fluid, and a typical compression
move is therefore very unlikely to succeed. On the other hand, by keeping
the chemical potential constant, the density can be more easily increased by
adding new particles. The transition calculated via the thermodynamic 
integration route
is completely consistent with our simulation data, and the transition is
predicted to fall inside the observed ``hysteresis'' loop.

We have verified the computed ($\mu $,$P$%
)-curves of the fluid, by calculating independently the chemical potential,
using the Widom particle-insertion method\cite{Widom},
in an NVT-simulation. 
%

\subsection{Gibbs Ensemble Simulations}

A method that focuses specifically on the location of the liquid-vapor
coexistence curve is the Gibbs-Ensemble technique\cite{Panagiotopoulos}.
Here two simulations are carried out in parallel: one in the liquid phase
and one in the vapor. The two systems are held at the same temperature and
are allowed to exchange volume and mass, but the total volume and total
number of particles of the two systems is fixed. This ensures that, at
equilibrium, the pressure and chemical potential of the two systems are the
same. As a consequence, the conditions for phase coexistence are
automatically satisfied. Using this technique we calculated liquid-gas
coexisting densities for the case of long-range attractions ($\alpha =0.1$),
where the liquid-vapor transition is stable and the phase diagram shows a
critical point as well as a triple point. The coexistence data have been
collected in Table \ref{LiqGas}. 
Close to the critical point the free energy
associated with the formation of the liquid-vapor interface becomes very
small. As a consequence, the free energy cost to create an interface in
either box becomes small, while the formation of such interfaces is
entropically favorable. For this reason, just below the critical point,
vapor-liquid coexistence can no longer be observed in a Gibbs Ensemble
simulation\cite{Smit}. Therefore the highest temperature at which the
coexistence can be observed is not a proper estimate of the critical
temperature of the system; nevertheless it is possible to estimate it by
assuming that the temperature dependence of the density difference of the two
coexisting phases can be fitted to a scaling law\cite{Rowlinson}: 
\begin{equation}
\rho _{liq}-\rho _{gas}=A\left( T-T_{c}\right) ^{\gamma }\qquad ,
\label{Scaling}
\end{equation}
where $\gamma $ is the critical exponent (for two dimensional systems $%
\gamma =0.125$), $T_{c}$ is the estimate of the critical temperature, and $A$
is a constant determined in the fit. Once $T_{c}$ is known, it is possible
to estimate the critical density $\rho _{c}$, by using the law of
rectilinear diameters\cite{Rowlinson}: 
\begin{equation}
\frac{\rho _{liq}+\rho _{gas}}{2}=\rho _{c}+B\left( T-T_{c}\right) \qquad ,
\label{Rectilinear}
\end{equation}
where $B$ is an adjustable parameter.

\section{Results}

\begin{figure}[tbp]
\center{\begin{minipage}{6.7cm}
    \epsfxsize 6.0 cm
    \center{\rotatebox{-90}{\epsfbox{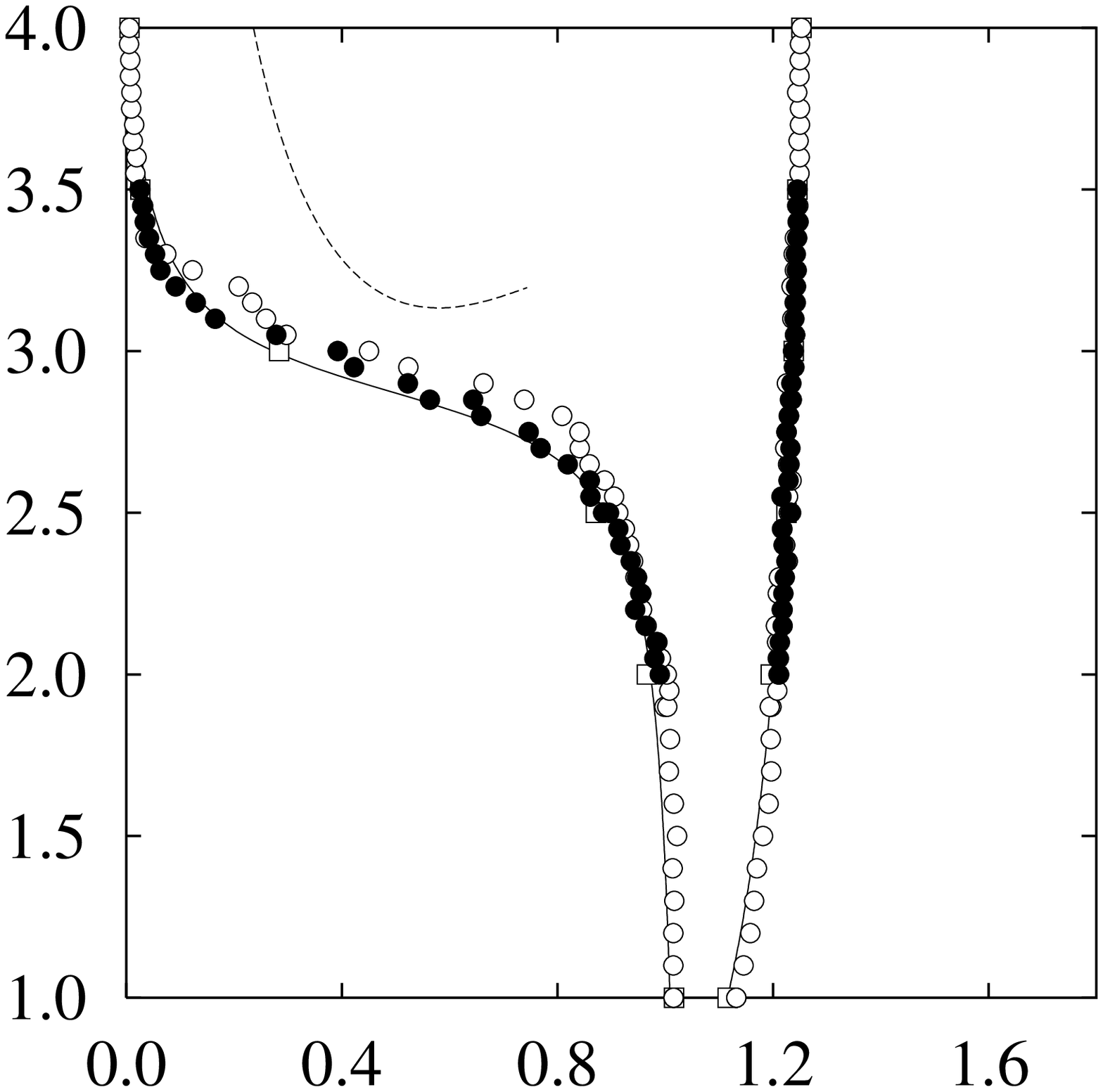}}}
    \center{(a) $\alpha=50$ short-range attr.}
  \end{minipage}
\put(-7.0, 2.5){$\beta$} 
\put(-0.2,-2.5){$\rho \sigma_{eff}^3$}}
\par
\center{\begin{minipage}{6.7cm}
    \epsfxsize 6.0 cm
    \rotatebox{-90}{\epsfbox{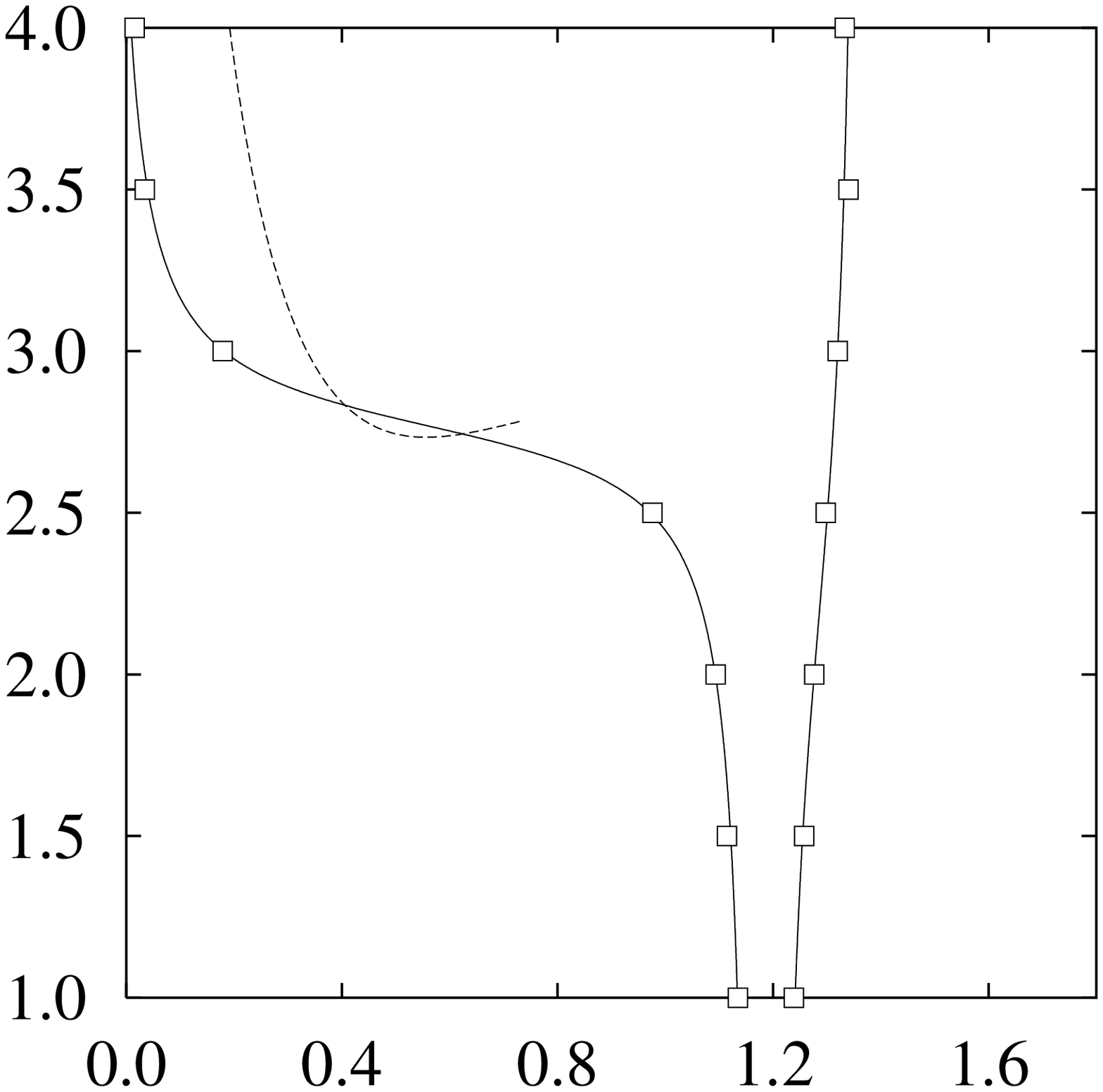}} 
    \center{(b) $\alpha=4.5$ intermediate-range attr.}
  \end{minipage}
\put(-7.0,3.0){$\beta$} 
\put(-0.2,-2.0){$\rho \sigma_{eff}^3$}}
\par
\center{\begin{minipage}{6.7cm}
    \epsfxsize 6.0 cm
    \center{\rotatebox{-90}{\epsfbox{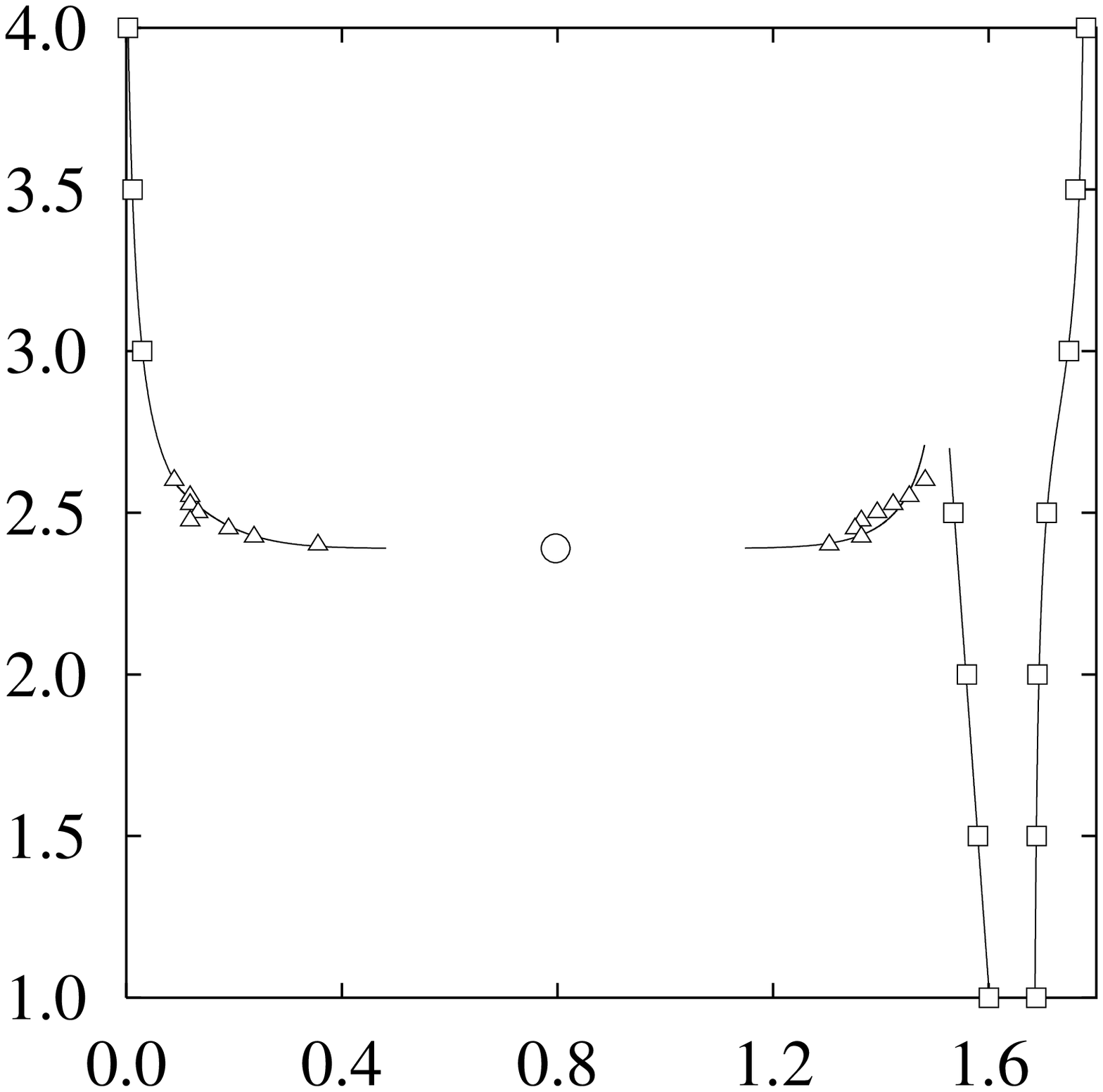}}} 
    \center{(c) $\alpha=0.1$ long-range attr.}
  \end{minipage}
\put(-7.0, 2.5){$\beta$} 
\put(-0.2,-2.5){$\rho \sigma_{eff}^3$}}
\par
\caption{Phase diagrams in the $(\protect\beta,\protect\rho )$
representation, shown for increasing range of attractions:
(a) $\alpha=50$, (b) $\alpha=4.5$, and (c) $\alpha=0.1$. 
The circles refer
to the Gibbs-Duhem integration method. Here we show 
(open circles)
a series for increasing $\protect%
\beta$ (decreasing temperature) and 
(filled circles)
a series for decreasing $\protect%
\beta$. The squares 
(open squares)
are the result of thermodynamic
integration, and the line just a guide to the eye. The fluid-fluid
equilibrium densities are calculated with Gibbs Ensemble simulations 
(open triangles)
when
possible, otherwise we estimated the spinodals 
(dashed line)
extrapolating the isothermal
compressibility. See text for details.}
\label{Beta-rho}
\end{figure}

We have mapped the phase boundaries for three different ranges of
attractions corresponding to $\alpha =50,4.5$ and $0.1$. Our results are
summarized in Figure (\ref{Beta-rho}). The first point to note is that we
find clear evidence of a {\it first order} transition between the solid and
fluid phase at finite temperatures. This finding would be trivial in three
dimensions but not in two. In fact there is considerable evidence that
melting in two dimensions may be a continuous phase transition\cite
{experiments,theory}. On the other hand evidence for first order
two-dimensional melting has also been observed in a number of systems\cite
{Gelbart-BenShaul} and there is, in fact, no theoretical reason to rule out
first-order melting in two dimensions.
%

\subsection{Short-range}

In the first panel of Figure (\ref{Beta-rho}) we show the calculated phase
diagram for the shortest range considered. A relatively small system size of 
$N=256$ particles was used in the mapping of the phase boundary using two
parallel NPT-simulations combined into the Gibbs-Duhem method. In order to
shorten simulation times we have truncated, but not shifted, the potential
at $r=3.5\sigma $, and maintained a neighbor list of particles within a
radius of $r=5.0\sigma $. In order to prevent the solid system from melting
and the liquid from crystallizing, we have used the artificial constraint on
a crystallinity order parameter (see Paragraph \ref{Method}). However, we
also ran a few simulations without this artificial constraint and we verified
that the phase coexistence was not an artifact due to the constraint.  We
performed two sets of integrations, one where we increased $\beta $ step-wise
(empty circles in the figure), and one in the opposite sense (filled
circles). An independent evaluation of the crystallization boundary was
obtained with thermodynamic integration. The squares in the figure represent
the calculated coexisting densities, and are connected with a simple fitting
function which is only meant to be a guide to the eye. Note that the curves
for increasing and decreasing $\beta $ do not superimpose everywhere. This
is due to the limited numerical accuracy of our Gibbs-Duhem integration.

It is interesting to compare our results with the phase diagram calculated
for particles interacting through the same potential and with same range\cite
{PR}, but in {\it three} dimensions. Constraining the system to two
dimensions causes the ``shoulder'' in the crystallization curve  to become
flatter and to move to  lower temperatures. The latter effect is not
surprising as there are more neighbors in $d=3$ than in $d=2$. For instance,
a sphere in a $d=2$ close-packed structure has 6 neighbors and in $d=3$, 12
neighbors. All other things being equal, in three dimensions the freezing
temperature is raised by a factor proportional to the number of neighbors.
Next, the effect of thermal fluctuations increases as the dimensionality of
the system is reduced. In $d=1$, solids can not exist because of thermal
fluctuations. In $d=2$, there is no true long-range positional order,
although there is a phase transition separating the liquid and the
crystalline phase.   We should expect that, due to the stronger fluctuations
in two dimensional fluids,  the (meta-stable) liquid-vapor critical
temperature, $T_{c}$, should be reduced compared to the corresponding
three-dimensional model. 

We have attempted to locate the metastable fluid-fluid equilibrium by
performing Gibbs-Ensemble simulations in the region where we estimated
demixing of the metastable fluid phase to occur. However, in these
simulations we saw no fluid-fluid demixing. Rather, we noticed a strong
tendency towards spontaneous crystallization. This behavior is in contrast with
analogous calculations in the corresponding three-dimensional system\cite{PR}%
. We argue that in two dimensions the local hexagonal structure of the dense
fluid, is similar to that of the solid. Only small density fluctuations
are necessary to overcome a (presumably) small free energy barrier for the
formation of the critical nucleus. Nevertheless we can still provide an
estimate of the fluid-fluid metastable equilibrium by extrapolating to the
temperature where the inverse isothermal compressibility, $\kappa
^{-1}=-V\left( \partial P/\partial V\right) _{T}$, of the  liquid phase
vanishes. This is straightforward, since the equation of state of the liquid
is known for several temperatures from the thermodynamic integration
procedures. See for example Figure (\ref{P-rho}). The inverse isothermal
compressibility can be rewritten as: 
\begin{equation}
\beta \kappa ^{-1}=\rho \left( \frac{\partial \beta P}{\partial \rho }%
\right) _{T}\qquad .  \label{Compress}
\end{equation}
We assumed that the inverse compressibility depends linearly on temperature
(this is not true close to the critical point, but there we have no data
anyway).The set of points where the extrapolated $\beta \kappa ^{-1}$
vanishes, provides us with our estimate of the meta-stable liquid-vapor
spinodal. This estimate is shown as a dotted curve in Figure (\ref{Beta-rho}%
).

\subsection{Intermediate-range}

We used thermodynamic integration to map out the solid-liquid equilibrium
for longer range of attractions. Even though the calculation is limited to a
few selected temperatures, this method has the advantage that it is more
robust than the Gibbs-Duhem. The curve connecting the squares in Figure (\ref
{Beta-rho}) is only a guide to the eye. As we increase the range of
attractions, we expect the metastable critical point to move to higher
temperatures. Indeed our estimate of the critical point, obtained using the
extrapolation method described above, predicts that, for the intermediate
range  ($\alpha =4.5$), the spinodal just about touches the crystallization
boundary. We attempted to perform Gibbs ensemble simulations to study the
liquid-vapor transition in this model system, but again we encountered a
strong tendency of the dense phase to crystallize.

\subsection{Long-range}

The situation changes quite dramatically when the range of attraction is
increased even more. The liquid-gas transition becomes stable and
Gibbs-Ensemble simulations are successful in locating the phase boundaries.
In Table \ref{LiqGas} we collect the coexistence densities as a function of
the inverse temperature. The uncertainties $\Delta \rho$ quoted in the table
are not the errors in the average densities, rather they refer to the
half-width of the histogram indicating the probability of finding a certain
density during the simulation. By fitting the liquid and gas densities to
the law of rectilinear diameters, we extrapolate the critical point at $%
T_c=0.418$ and $\rho_c=0.134$. The fitting curve is portrayed in the third
panel of Figure (\ref{Beta-rho}). The open circles are the Gibbs-Ensemble
simulation coexistence densities for the liquid-gas equilibrium, while the
open squares refer to the solid-fluid equilibrium and were determined
through thermodynamic integration. At very low temperatures, though,
calculating the fluid branch of the equation of state becomes quite a
difficult task, because equilibration times become increasingly longer. Here
we have estimated the crystallization boundary by Grand-Canonical ($\mu$VT)
simulations of very low density systems.
%
\begin{figure}[tbp]
\centering{
\begin{minipage}{6.7cm}
    \epsfxsize 6.0 cm
    \rotatebox{-90}{\epsfbox{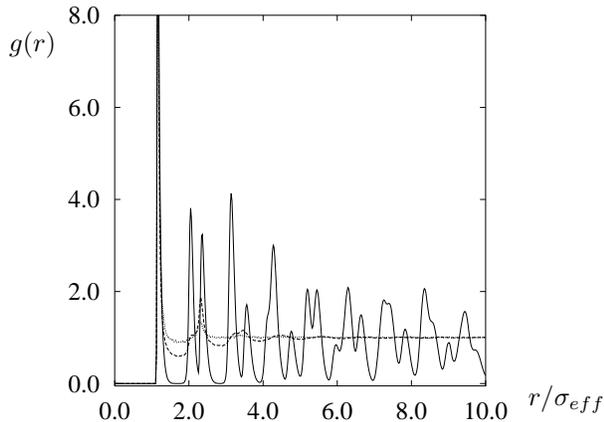}}
  \end{minipage}
\put(-7.1, 2.2){$g(r)$} 
\put(-0.2,-2.5){$r/\sigma_{eff}$}}
\par
\caption{Pair correlation function $g(r)$ calculated in the short-range
attraction $\protect\alpha =50$ system at $T^*=2.5$ for different densities:
(dotted line)
typical
gas density $\protect\rho^*=0.13$, 
(dashed line)
liquid system exactly at
equilibrium $\protect\rho^*=0.5680$, and 
(solid line)
the corresponding coexisting solid
system $\protect\rho^*=0.8267$.}
\label{gr}
\end{figure}

\subsection{Structural properties}

\begin{figure}[tbp]
\centering{
\begin{minipage}{6.7cm}
    \epsfxsize 6.0 cm
    \rotatebox{-90}{\epsfbox{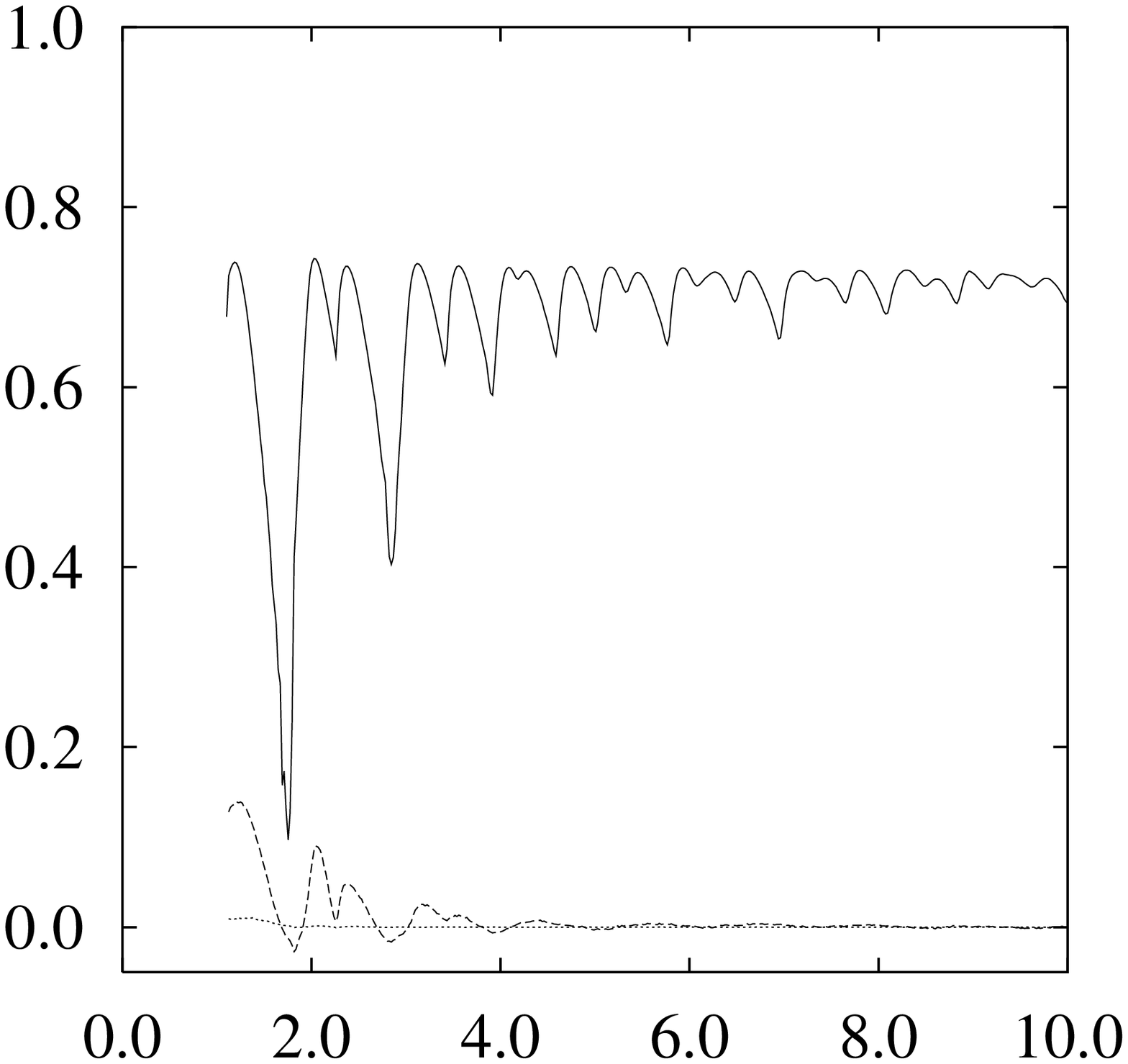}}
  \end{minipage}
\put(-7.1, 2.2){$g_6(r)$} 
\put(-0.2,-2.5){$r/\sigma_{eff}$}}
\par
\caption{The bond orientational correlation function $g_6(r)$ calculated for
the systems described in Fig. (\ref{gr}). Note how the low density system
carries no bond correlations even at short distances; the correlation in the
liquid is lost within $4-5$ particle diameters, but in the solid the
bond-order parameter tends to a non-zero value in the limit of long
particle-particle separation.}
\label{g6}
\end{figure}

In order to investigate the structural properties of the phases identified
in the previous sections, we have performed simulations on a larger system
of $450$ particles, at a single temperature, for the case of a short-range
potential. The use of this larger system enables the calculation of the pair
distribution functions for large separation: this is especially important in
order to study the decay of the bond-order parameter close to the solid-liquid
transition. Monte Carlo simulations were performed in the NPT-ensemble. 
The
structures of the two coexisting phases were characterized using the radial
distribution function $g(r)$ (see Fig. (\ref{gr})) and the bond-order
correlation function $g_{6}(r)$, which correlates over the value of
the {\it local} bond-orientational order parameter $\psi _{6}$ between two
particles. This is defined by: 
\begin{equation}
g_{6}(r)=\frac{\langle \psi _{6}^{\ast }({\bf r}_{j})\psi _{6}({\bf r}%
_{i})\rangle }{g(r)}\qquad ,  \label{gr6}
\end{equation}
where the local bond orientational order for particle $i$ at a position $%
{\bf r}_{i}$ is given by 
\begin{equation}
\psi _{6}({\bf r}_{i})=\frac{\sum_{k}w(r_{ik})\exp (6i\theta _{ik}))}{%
\sum_{k}w(r_{ik})}\qquad .  \label{phi6}
\end{equation}
In this expression the summation is over the neighboring particles $k$ of
particle $i$ and $\theta _{ik}$ is the angle between the vector $({\bf r}%
_{i}-{\bf r}_{k})$ and a fixed reference axis.  We used a weighting function 
$w$ to define nearest neighbors\cite{Strandberg}. In the present study, 
we chose $w(r)$ such that it is unity for a separation of $r<1.6\sigma $ and
zero for $r_{ik}$ above $1.8\sigma $ with a linear interpolation between the
two endpoints. The upper limit of the weighting function was chosen such
that all the particles included in the first peak of the pair distribution
function contribute to $\psi _{6}$. The bond-order correlation function
is large if local bond-order  parameters are correlated over large distances
(as they are in the crystalline solid). In the isotropic liquid, bond-order
correlations decay exponentially. However, in the hexatic phase, an algebraic
decay is expected.  Our results for $g_{6}(r)$ are shown in Figure (\ref{g6}%
): it is apparent that $g_{6}(r)$ rapidly tends to zero in the liquid phase,
and therefore the liquid phase at coexistence is not even close to becoming
hexatic.  Of course, in the coexisting solid phase, the bond orientational
correlation function tends to a non-zero value, as it should. These findings
support our ``thermodynamic'' observation that the fluid-solid transition in
these model systems appears to be first order.
%

\section{Conclusions}

We have studied the phase behavior of a simple model system that is meant to
mimic membrane proteins confined in a quasi-two-dimensional geometry. To
study the phase behavior, we used a variety of complementary simulation
techniques. Where possible we have used various routes to estimate the phase
boundaries. Our study focused on the influence of the range of the
attractive interactions on the topology of the phase diagram.  For
long-ranged attractions the phase diagram displays a stable liquid-vapor
critical point and a solid-liquid-vapor triple point. As the range of
attraction is decreased,  the stable liquid-gas transition becomes
metastable and the critical point moves into the solid-fluid two-phase
region.  Qualitatively, the phase diagrams of the two-dimensional systems
that we studied are similar to those of their three-dimensional
counterparts. However, quantitatively, there are large differences. Most
importantly, we find that in systems where the liquid-vapor coexistence
curve has moved below the freezing curve, there is virtually no barrier to
crystallization. This suggests that membrane proteins with effectively
isotropic interactions should easily form two-dimensional crystals. Two
questions arise: 1.) how do these two dimensional crystals proceed to form
three-dimensional crystals and 2.) to what extent is the ease of  $2d$
crystallization changed  by anisotropy in the protein-protein interactions?

\section{Acknowledgments}

The work of the FOM Institute is part of the research program of ``Stichting
Fundamenteel Onderzoek der Materie'' (FOM) and is supported by NWO.
Computing time was provided by SARA (Stichting Academisch Rekencentrum
Amsterdam). The authors gratefully acknowledge Martin Bates for useful discussions
and for helping with the details of the simulations. The authors also thank 
Anand Yethiraj and Jurgen Horbach for critical reading of the manuscript. 
MGN acknowledges
financial support from EU contract ERBFMBICT982949.

%
\begin{table}[tbp]
\begin{tabular}{lllll}
$\beta$ & $\rho_{gas}$ & $\Delta \rho$ & $\rho_{liq}$ & $\Delta \rho$ \\ 
\hline
2.60 & 0.015 & 0.010 & 0.250 & 0.005 \\ 
2.55 & 0.020 & 0.010 & 0.245 & 0.005 \\ 
2.525 & 0.020 & 0.010 & 0.240 & 0.010 \\ 
2.50 & 0.0225 & 0.010 & 0.235 & 0.010 \\ 
2.475 & 0.020 & 0.015 & 0.230 & 0.015 \\ 
2.45 & 0.032 & 0.015 & 0.228 & 0.010 \\ 
2.425 & 0.040 & 0.020 & 0.230 & 0.020 \\ 
2.40 & 0.060 & 0.015 & 0.220 & 0.015 \\ 
2.375 & 0.070 & 0.020 & 0.200 & 0.020 \\ 
2.35 & 0.090 & 0.020 & 0.180 & 0.020
\end{tabular}
\caption{Coexistence data for the liquid-gas equilibrium. $\protect\beta$ is
the inverse temperature and $\protect\rho$ is the number density. The
uncertainties $\Delta \protect\rho$ quoted here refer to the half-width of
the histogram indicating the probability of finding a certain density during
the simulation. By fitting $\protect\rho_{gas}$ and $\protect\rho_{liq}$ to
the law of rectilinear diameters, we extrapolate the critical point at $%
\protect\beta_c=2.392$ and $\protect\rho_c=0.134$.}
\label{LiqGas}
\end{table}

\end{document}